\documentclass[final]{iopart}
\usepackage{iopams,graphicx,texdraw,epsf,bm,texdraw}
\begin{document}
\paper[Boundary critical behaviour at $m$-axial Lifshitz
points]{Boundary critical behaviour at $\bm{m}$-axial Lifshitz points
  of semi-infinite systems with a surface plane perpendicular to a
  modulation axis}

\author{H.~W. Diehl,$^{1,2}$ M.~A. Shpot$^{1,3}$ and P.~V. Prudnikov$^{1,4}$}
\address{$^1$ Fachbereich Physik, Universit{\"a}t
Duisburg-Essen, Campus Essen, D-45117 Essen, Federal Republic of
Germany}
\address{$^2$ Department of Physics and Center for Stochastic
  Processes in Science and Engineering, Virginia Tech, Blacksburg,
  Virginia 24061-0435, USA}
\address{$^3$ Institute for Condensed Matter Physics, 79011 Lviv,
  Ukraine}
\address{$^4$ Department of Theoretical Physics, Omsk State University,
Mira Prospect 55-A, Omsk, 644077, Russia}
\begin{abstract}
  Semi-infinite $d$-dimensional systems with an $m$-axial bulk
  Lifshitz point are considered whose ($d-1$)-dimensional surface
  hyper-plane is oriented perpendicular to one of the $m$ modulation
  axes. An $n$-component $\phi^4$ field theory describing the bulk and
  boundary critical behaviour when (i) the Hamiltonian can be taken to
  have $O(n)$ symmetry and (ii) spatial anisotropies breaking its
  Euclidean symmetry in the $m$-dimensional coordinate subspace of
  potential modulation directions may be ignored is investigated. The
  long-distance behaviour at the ordinary surface transition is mapped
  onto a field theory with the boundary conditions that both the order
  parameter $\bm{\phi}$ and its normal derivative
  $\partial_n\bm{\phi}$ vanish at the surface plane. The
  boundary-operator expansion is utilized to study the short-distance
  behaviour of $\bm{\phi}$ near the surface. Its leading contribution
  is found to be controlled by the boundary operator
  $\partial_n^2\bm{\phi}$.  The field theory is renormalized for
  dimensions $d$ below the upper critical dimension $d^*(m)=4+m/2$,
  with a corresponding surface source term $\propto
  \partial_n^2\bm{\phi}$ added. The anomalous dimension of this
  boundary operator is computed to first order in $\epsilon=d^*-d$.
  The result is used in conjunction with scaling laws to estimate the
  value of the single independent surface critical exponent
  $\beta_{\mathrm{L}1}^{(\mathrm{ord},\perp)}$ for $d=3$. Our estimate
  for the case $m=n=1$ of a uniaxial Lifshitz point in Ising systems
  is in reasonable agreement with published Monte Carlo results.
\end{abstract}
\ead{shpot@ph.icmp.lviv.ua}

\section{Introduction}
\label{sec:intro}

The significance of the $n$-component $\phi^4$ models with $O(n)$
symmetric Hamiltonian derives from the fact that they represent the
most common---and probably also most important---universality classes
of critical behaviour at bulk critical points of $d$-dimensional
systems with short-range interactions. Prominent examples of such
universality classes for given $d$ and $n=1$, $2$, and $3$ are those
of the Ising, XY, and isotropic Heisenberg models, respectively.

When such systems are bounded by free $(d-1)$-dimensional
hyper-surfaces or walls, a wealth of distinct boundary critical
phenomena can occur \cite{Bin83,Die86a,Die97,Ple04}. A well-studied
case are systems with a free surface that can be described by a
semi-infinite $n$-component $\phi^4$~model. The Hamiltonian of the
latter is of the form
\begin{equation}
  \label{eq:Ham}
    {\mathcal{H}}={\int_{\mathfrak{V}}}{\mathcal L}_{\rm b}(\boldsymbol{x})\,
    {\rm d} V+ {\int_{\mathfrak{B}}}{\mathcal L}_1(\boldsymbol{r})
    \,{{\rm d}}A\;, 
\end{equation}
where $\int_{\mathfrak{V}}$ and ${\int_{\mathfrak{B}}}$ mean volume
and surface integrals over the half-space
$\mathfrak{V}={\mathbb{R}}^d_+\equiv\{\bm{x}=(\bm{r},z)
|\bm{r}\in{\mathbb{R}}^{d-1},0\leq z<\infty\} $ and the $z=0$ boundary
hyperplane ${\mathfrak{B}}$, respectively. Provided neither bulk nor
surface terms breaking the $O(n)$~symmetry must be included, the bulk
and surface densities are given by
\begin{equation}
  \label{eq:Lbm0}
  {\mathcal{L}}_b=\frac{1}{2}\,(\nabla\phi)^2
  +\frac{\mathring{\tau}}{2}\,\phi^2 +
\frac{\mathring{u}}{4!}\,|\bm{\phi}|^4
\end{equation}
and
\begin{equation}
  \label{eq:L1m0}
  {\mathcal{L}}_1=\frac{\mathring{c}}{2}\,\phi^2\;.
\end{equation}
Furthermore, one can distinguish three different types of surface
transitions that take place at the bulk critical temperature $T_c$
\cite{Bin83,Die86a}: They are called ordinary, special, and
extraordinary surface transitions and occur, depending on whether
$\mathring{c}$ is larger than, equal to, or smaller than a certain
critical value $\mathring{c}_{\rm sp}$.

The extraordinary transition is a transition from a surface-ordered,
bulk-disordered high-temperature phase to a low-temperature phase with
long-range order both at the surface and in the bulk. It---and hence
also the special transition---can occur only for those choices of $d$
and $n$ for which fluctuations do not destroy long-range order at the
surface at all nonzero temperatures $T$. This requires $d-1>1$ in the
scalar case $n=1$, and $d-1>2$ in the continuous-symmetry case $n>1$.

Suppose $n$ and $d$ are such that all three of these transitions can
occur. Then the critical behaviour that surface quantities exhibit at
any of these is representative of a separate surface universality
class, although their bulk critical behaviour is the same. Thus the
bulk universality class associated with this choice of $n$ and $d$
splits into three distinct surface universality classes.

In the present paper we shall be concerned with the surface critical
behaviour of $n$-vector systems at $m$-axial bulk Lifshitz points. A
Lifshitz point (LP) is a multicritical point at which a disordered, a
homogeneous ordered, and a modulated ordered phase meet
\cite{Hor80,Sel92,Die02}. A family of natural extensions of the bulk
models defined on $\mathbb{R}^d$ by the bulk density (\ref{eq:Lbm0})
that have such LPs was introduced decades ago \cite{HLS75a,SG78}, but
investigated in greater detail via field-theoretic renormalization
group (RG) only in the past few years (see the
papers~\cite{MC98,MC99,DS00a,SD01} and their references). Their bulk
density is given by
\begin{equation}
  \label{eq:Lb}\fl
 {\mathcal L}_{\rm b}(\boldsymbol{x})=\frac{\mathring{\sigma}}{2}
  \Big(\sum_{\alpha=1}^m\partial_\alpha^2\boldsymbol{\phi} \Big)^2
 +\frac{1}{2}
\sum_{\beta=m+1}^d{(\partial_\beta\boldsymbol{\phi})}^2
+\frac{\mathring{u}}{4!}\,|\boldsymbol{\phi}|^4
+\frac{\mathring{\rho}}{2}
\sum_{\alpha=1}^m{(\partial_\alpha\boldsymbol{\phi})}^2
 +\frac{\mathring{\tau}}{2}\,
\boldsymbol{\phi}^2\;,
\end{equation}
where the position vector $\bm{x}$ has the representation
$(x_\gamma)=(x_\alpha,x_\beta)$ in Euclidean coordinates $x_\gamma$.
We use the convention that $\alpha$ and $\beta$ denote coordinate
indices $\gamma$ with $1\leq \alpha\leq m$ and $m<\beta\leq d$,
respectively.  Likewise, $\partial_\alpha$ and $\partial_\beta$ mean
the corresponding spatial derivatives
$\partial_\gamma=\partial/\partial x_\gamma$. At the level of Landau
theory, the model has a continuous transition from a disordered to a
homogeneous ordered phase for $\mathring{\rho}>0$ at
$\mathring{\tau}=0$. For negative $\mathring{\rho}$, a continuous
transition to a modulated ordered phase occurs at a nonzero value of
$\mathring{\tau}$. The transition lines between these phases merge at
$\mathring{\tau}=\mathring{\rho}=0$, which is an $m$-axial LP within
Landau theory.

At LPs, systems exhibit scale invariance of the \emph{strong anisotropic
kind}: coordinate difference $\Delta x_\alpha$ within the
$\alpha$-subspace scale as a nontrivial power $(\Delta
x_\beta)^\theta$ of the complementary ones $\Delta x_\beta$, where
$\theta$, the anisotropy exponent, generally differs from $1$, and in
Landau theory has the mean-field value $\theta^{\rm MF}=1/2$. 

Owing to this anisotropic scale invariance, boundary critical
phenomena at LPs are richer than at critical points (CPs). For
example, in the CP case, the orientation of the surface relative to
the coordinate axes does not play any important role on the level of a
description in terms of a $\phi^4$ field theory. However, for systems
at LPs the surface's orientation matters in an essential way, just as
it does quite generally for systems with anisotropic scale invariance:
Two fundamentally distinct orientations can be distinguished---one for
which all $\alpha$-directions are parallel to the surface (which we
shall refer to as parallel), and another one for which the surface
normal $\bm{n}$ is along one of the $m$ $\alpha$-directions (which we
shall refer to as perpendicular).

As a consequence of the different scaling behaviour of distances along
$\alpha$- and $\beta$-directions, it depends on whether the
orientation of the surface is parallel or perpendicular which boundary
contributions ${\mathcal L}_1$ are potentially infrared relevant below
the upper critical dimension $d^*(m)=4+m/2$ and hence must be included
in the action. The problem of constructing semi-infinite extensions
of the models with bulk density~(\ref{eq:Lb}) for $d=d^*(m)-\epsilon$
that are \emph{``minimal''} in the sense that all irrelevant and marginal
boundary contributions not compatible with the presumed $O(n)$ and
Euclidean%
\footnote{We presume that bulk and boundary contributions breaking the
  rotational invariance in the $\alpha$-subspace may be ignored.  When
  $m>1$, spatial anisotropies lead, in particular, to non-isotropic
  bulk contributions of the form
  $T_{\alpha_1\alpha_2\alpha_3\alpha_4}\,
  (\partial_{\alpha_1}\partial_{\alpha_2}\bm{\phi})
  \cdot\partial_{\alpha_3}\partial_{\alpha_4}\bm{\phi}$, where
  $T_{\alpha_1\alpha_2\alpha_3\alpha_4}$ is a tensor whose symmetry is
  that of an $m$-cube or lower. The $O(\epsilon^2)$~results of
  reference~\cite{DSZ03} for the bulk model indicate that such terms
  are relevant in the infrared for $\epsilon>0$.}
symmetries are discarded was considered for the case of parallel
surface orientation in references~\cite{DRG03,DGR03}, where it was
found that a contribution
$\propto\sum_\alpha(\partial_\alpha\bm{\phi})^2$ had to be included in
the corresponding surface density ${\mathcal L}_1\equiv {\mathcal
  L}_1^\|$, in addition to the one in equation~(\ref{eq:L1m0}). The
so-obtained semi-infinite model, defined by equation~(\ref{eq:Lb}) in
conjunction with the boundary density
\begin{equation}
  \label{eq:L1par}
  {\mathcal L}_1^\|=\frac{\mathring{c}}{2}\,\phi^2
  +\frac{\mathring{\lambda}}{2}\,
  \sum_{\alpha=1}^m(\partial_\alpha\bm{\phi})^2\;,
\end{equation}
was then utilized in references~\cite{DGR03} and \cite{DR04} to
determine the $\epsilon$~expansion of the surface critical exponents of the
corresponding ordinary and special transitions to second and first
order, respectively. This extends or complements previous work based
on the mean-field approximation \cite{FKB00} and Monte Carlo
simulations for the axial-next-nearest-neighbour Ising (ANNNI) model
\cite{Ple02}.

The model is a natural and simple-looking generalization of the
semi-infinite $n$-vector model defined by equations~(\ref{eq:Lbm0}) and
(\ref{eq:L1m0}), to which it reduces when $m=0$. The fluctuating
boundary conditions \cite{Die86a,Die97} one obtains for it,
\begin{equation}
  \label{eq:bcparcase}
  \partial_n\bm{\phi}=(\mathring{c}
  -\mathring{\lambda}\,\partial_\alpha\partial_\alpha)\, \bm{\phi}\;,
\end{equation}
where $\partial_n$ ($=\partial_z$) is the derivative along the inward
normal $\bm{n}$, are of the Robin type, generalizing the familiar ones
of the CP ($m=0$) case to nonzero values of $\mathring{\lambda}$.%
\footnote{Here and below we utilize the summation convention that
  pairs of the same $\alpha$- and $\beta$-indices are to be summed
  from $\alpha=1$ to $m$ and $\beta=m+1$ to $d$, respectively.}
As shown in references~\cite{DGR03,DR04}, the new boundary term $\propto
\mathring{\lambda}$ has the effect of moving the fixed points onto
which the ordinary, special, and extraordinary transitions are mapped
to a nontrivial value $\lambda^*=O(\epsilon^0)$ of the renormalized
counterpart $\lambda$ of $\mathring{\lambda}$.

In the following we shall consider the case of \emph{perpendicular}
surface orientation. Previous work on it either was restricted to an
investigation of the corresponding ordinary transition of the
semi-infinite ANNNI model on the level of the mean-field approximation
\cite{Gum86,BF99}, or else employed Monte Carlo simulations to study
both the ordinary and special transitions of this model \cite{Ple02}.

Our aim is to construct and analyze an appropriate minimal
semi-infinite model for general values of $n$, $m$ and $d$. This turns
out to be a greater challenge and more interesting than in the case of
parallel surface orientation. Since the classical equation of motion
of the order-parameter profile now involves a differential equation of
fourth order in $\partial_z$, two rather than one boundary condition
are required (aside from two analogous ones at $z=\infty$). Furthermore,
being an $x_\alpha$ coordinate, $z\equiv x_m$ now scales naively as
$\mu^{-1/2}$ where $\mu$ is an arbitrary momentum unit making $\mu
x_\beta$ dimensionless. Compared to the case of parallel surface
orientation (where $z\sim \mu^{-1}$, power counting gives more
potentially relevant or marginal surface terms. Recently one of us has
suggested an appropriate surface density $\mathcal{L}_1^\perp$ and the
associated boundary conditions \cite{Die05}. In
section~\ref{sec:model}, we recall this density and derive the
fluctuating boundary conditions it implies. Using the latter, we show
that the existing additional surface contributions that are
compatible with symmetry and short-rangedness of interactions, and not
irrelevant according to power counting, are redundant.

In section~\ref{sec:freeprop} we determine the free propagator subject
to these boundary conditions, at the LP for general values of the surface
interaction constants. Owing to its rather complicated form, explicit
perturbative RG calculation for dimensions $d={d^*(m)-\epsilon}$ are
difficult to perform with it. In section~\ref{sec:BOT} we show,
following a suggestion made in reference \cite{Die05}, how this
problem can be by-passed to some extent in the case of the ordinary
transition. The asymptotic large-scale behaviour at this transition can
be argued to be described by a theory with boundary conditions
\begin{equation}
  \label{eq:bcperp}
  \phi=\partial_n\phi=0\;.
\end{equation}
Hence one can work with the corresponding simplified free propagator
at the price of having to deal with correlation functions involving
besides the field $\bm{\phi}$ also the boundary operator
$\partial_n^2\bm{\phi}$.  From the anomalous dimension of the latter
the required single independent surface critical exponent
$\beta_{\mathrm{L}1}^{(\mathrm{ord},\perp)}$ of this transition can be
inferred. Performing a one-loop RG analysis, we compute in
section~\ref{sec:RG} the $\epsilon$~expansion of this exponent to
$\Or(\epsilon)$ for general values of $m$. The result is used to
estimate its value for the case $m=n=1$ of the three-dimensional
semi-infinite ANNNI model.  Section~\ref{sec:concl} contains
concluding remarks. Finally, there is an appendix explaining how the
required one-loop integral was calculated.

While it is quite common to investigate field theories with Dirichlet
boundary conditions (see \cite{Die86a,Die97,KD92a,MO93a} and their
references), we are not aware of previous studies of field theories
satisfying the boundary conditions~(\ref{eq:bcperp}), barring familiar
examples of hydrodynamic equations for fluids \cite{Cha70}. It
therefore not unlikely that the work described below might also be of
interest for other problems.

\section{Boundary density and fluctuating boundary conditions}
\label{sec:model}

The boundary density suggested in reference~\cite{Die05} is
\begin{equation}
  \label{eq:L1perp}\fl
  {\mathcal L}^\perp_1=\frac{\mathring{c}_\perp}{2}\,\phi^2
+\mathring{b}\,\boldsymbol{\phi}\partial_n\boldsymbol{\phi}
  +\sum_{\alpha=1}^{m-1}\Big[ \frac{\mathring{\lambda}_\|}{2}\,
  (\partial_\alpha\boldsymbol{\phi})^2 +\mathring{f}\,
  (\partial_\alpha\boldsymbol{\phi})
  \partial_n\partial_\alpha\boldsymbol{\phi}\Big]
  +\frac{\mathring{\lambda}_\perp}{2}\,(\partial_n\boldsymbol{\phi})^2
\end{equation}
with $\mathring{f}=0$. We have added the term $\propto \mathring{f}$
since it cannot be ruled out. Contributions breaking the symmetry
among the $m-1$ $\alpha$-directions parallel to the surface have been
excluded. That different values should be allowed for the coupling
constants $\mathring{\lambda}_\perp$ and $\mathring{\lambda}_\|$
should be obvious because the surface breaks the symmetry between
$\alpha$-directions parallel to it and the $z$-direction. Besides the
monomials included in equation~(\ref{eq:L1perp}), there are a number
of other surface operators one has to worry about, namely
\begin{equation}
  \label{eq:monomials} 
  \mathcal{O}_1=\bm{\phi}\partial_n^2\bm{\phi}\;,\quad
 \mathcal{O}_2=(\partial_n\bm{\phi})\partial_n^2\bm{\phi}\;,\quad
 \mathcal{O}_3=\bm{\phi}\partial_n^3\bm{\phi}\;.
\end{equation}
Before discussing these, let us first derive the fluctuating boundary
conditions that apply to the model defined by
equations~(\ref{eq:Ham}), (\ref{eq:Lb}) and (\ref{eq:L1perp}). 

To this end, we compute the variation $\delta\mathcal{H}$ of the
Hamiltonian. By integrations by parts, one gets
\begin{eqnarray}
  \label{eq:delH}
  \delta\mathcal{H}=\int_{\mathfrak{V}} \delta\bm{\phi}\left\{
    \frac{\partial\mathcal{L}_b}{\partial\bm{\phi}}  +
    \sum_{\alpha=1}^m\partial_\alpha^2 
  \frac{\partial{\mathcal L}_b}{\partial(\partial_\alpha^2\bm{\phi})}
  -\sum_{\gamma=1}^d\partial_\gamma\frac{\partial{\mathcal
      L}_b}{\partial(\partial_\gamma\bm{\phi})}\right\}\nonumber\\ 
  \lo\strut +{\int_{\mathfrak{B}}}\delta\bm{\phi}\left[
\frac{\partial{\mathcal
      L}^\perp_1}{\partial\bm{\phi}}-  \frac{\partial{\mathcal
      L}_b}{\partial(\partial_n\bm{\phi})}+
  \partial_n\frac{\partial{\mathcal L}_b}{\partial(\partial_n^2\bm{\phi})}-
  \sum_{\alpha=1}^{m-1} \partial_\alpha\frac{\partial{\mathcal
      L}^\perp_1}{\partial(\partial_\alpha\bm{\phi})}
\right]\nonumber\\
\lo\strut+{\int_{\mathfrak{B}}} (\partial_n\delta\bm{\phi})\left[
  \frac{\partial{\mathcal L}^\perp_1}{\partial(\partial_n\bm{\phi})}- 
  \frac{\partial{\mathcal L}_b}{\partial(\partial_n^2\bm{\phi})} -
  \sum_{\alpha=1}^{m-1} \partial_\alpha\frac{\partial\mathcal{ 
      L}^\perp_1}{\partial(\partial_n\partial_\alpha\bm{\phi})}\right].
\end{eqnarray}
Equating the expression in curly brackets to zero gives us the
classical field equation 
\begin{equation}
  \label{eq:beos}
  \left[\mathring{\sigma}(\partial_\alpha\partial_\alpha)^2
  -\mathring{\rho}\,\partial_\alpha\partial_\alpha
  -\partial_\beta\partial_\beta +\mathring{\tau}
  +\frac{\mathring{u}}{6}\,\phi^2\right]\bm{\phi}=\bm{0}\;.
\end{equation}
Doing the same with the expressions in square brackets of the surface
integrals ${\int_{\mathfrak{B}}}\delta\bm{\phi}[\ldots]$ and
${\int_{\mathfrak{B}}}\delta\partial_n\bm{\phi}[\ldots]$ yields the boundary
conditions
\begin{equation}
  \label{eq:bcperp1}
  {\Big\{\mathring{\sigma}\partial_n^3+(\mathring{b}-
  \mathring{\rho})\partial_n+\mathring{c}_\perp
  -[\mathring{\lambda}_\|+(\mathring{f}-\mathring{\sigma})\partial_n]
  \sum_{\alpha=1}^{m-1}\partial_\alpha^2\Big\}}\boldsymbol{\phi} =0
\end{equation}
and
\begin{equation}
  \label{eq:bcperp2}
\left\{-\mathring{\sigma}\,\partial_n^2+\mathring{\lambda}_\perp\partial_n
  +\mathring{b}-(\mathring{f}+\mathring{\sigma})
  \sum_{\alpha=1}^{m-1}\partial_\alpha^2\right\}\boldsymbol{\phi}=0
\end{equation}
respectively. 

These boundary conditions hold in Landau theory. Yet, they remain
valid inside of averages, for the same reason that the classical
equation~(\ref{eq:beos}) does so. To show this, one can make a shift
$\bm{\phi}\to\bm{\phi}+\bm{\Phi}$ in the functional integral defining
the generating functional $\mathcal{Z}[\bm{J}]\propto \int
\mathcal{D}\bm{\phi}\,\exp\left(-\mathcal{H}
  +\int_{\mathfrak{V}}\bm{J}\cdot\bm{\phi}\right)$ of multi-point
correlation functions $\langle\phi\ldots\phi\rangle$. For a
$\bm{\Phi}$ independent of $\bm{\phi}$, the functional measure
$\mathcal{\bm{\phi}}$ does not change, and one arrives at the equation
\begin{equation}
  \label{eq:avJ}
  \Big\langle\delta\mathcal{H}
  -\int_{\mathfrak{V}}\bm{J}\cdot\bm{\Phi}\Big\rangle_{\bm{J}}=0\;,
\end{equation}
where $\bm{\Phi}$ must be substituted for $\delta\bm{\phi}$ in
$\delta\mathcal{H}$. Here $\langle.\rangle_{\bm{J}}$ indicates a
normalized average in the presence of the source $\bm{J}$, i.e., with
the weight $\exp\left(-\mathcal{H}
  +\int_{\mathfrak{V}}\bm{J}\cdot\bm{\phi}\right)$. From this
result~(\ref{eq:avJ}) the validity of
equations~(\ref{eq:beos})--(\ref{eq:bcperp2}) inside of averages can
be derived in a well-known fashion by setting the source to zero,
either directly or after taking functional derivatives with respect to
it, and exploiting the arbitrariness of $\bm{\Phi}$ at and away from
the boundary.  As usual, the source term yields extra contributions
located at coinciding points to correlation functions generated by
functional differentiation of equation~(\ref{eq:avJ}). Corresponding
fluctuation corrections would result from additional surface source
terms such as $\int_{\mathfrak{B}}\bm{J}_1\cdot\bm{\phi}$.

It is now easy to understand why the surface operators given in
equation~(\ref{eq:monomials}) do not have to be included in the
Hamiltonian: These monomials involve $\partial_n^2\bm{\phi}$ and
$\partial_n^3\bm{\phi}$. We can solve the boundary
conditions~(\ref{eq:bcperp2}) and (\ref{eq:bcperp1}) for these
operators and substitute the solutions into the monomials. The
expressions that result for the surface operators $\mathcal{O}_1$,
$\mathcal{O}_2$ and $\mathcal{O}_3$ are linear combinations of the
surface operators retained in the surface density~(\ref{eq:L1perp}).%
\footnote{In expressions such as
  $\bm{\phi}\cdot\sum_{\alpha=1}^{m-1}\partial_\alpha^2\bm{\phi}$ and
  $\partial_n\bm{\phi}\cdot\sum_{\alpha=1}^{m-1}\partial_\alpha^2\bm{\phi}$,
  the derivatives $\partial_\alpha$ evidently can be made to act as
  $(-\overleftarrow{\partial}_\alpha)$ to
  the left, by means of integrations by parts.}
Thus the effect of the boundary operators $\mathcal{O}_1$,
$\mathcal{O}_2$ and $\mathcal{O}_3$ can be absorbed by a redefinition
of the surface variables of the surface density~(\ref{eq:L1perp}).
That is, they are redundant and may be discarded.

\section{Free propagator at the Lifshitz point}
\label{sec:freeprop}

We now turn to the calculation of the free propagator
$G(\bm{x},\bm{x}')$ at the LP. To this end we split the position
vector into components parallel and perpendicular to the surface,
writing $\bm{x}=(\bm{r},z)$. We choose $z$ to be $x_m$. The component
$\bm{r}$ then involves the $m-1$ $\alpha$-components
$(x_1,\ldots,x_{m-1})\equiv\bm{r}_<$ and the $d-m$ $\beta$-coordinates
$(x_\beta)=(x_{m+1},\ldots,x_d)\equiv\bm{r}_>$. For the wave-vector
conjugate to $\bm{r}=(\bm{r}_<,\bm{r}_>)$ we use analogous
conventions, writing
$\bm{p}=(\bm{p}_<,\bm{p}_>)$. 

From equation~(\ref{eq:beos}) one easily concludes that the partial
Fourier transform $\hat{G}(\bm{p};z,z')$, defined by
\begin{equation}
  \label{eq:Ghatdef}
  G(\bm{x},\bm{x}')=\int_{\bm{p}}\,
  \hat{G}(\bm{p};z,z')
  \,\mathrm{e}^{\rmi\bm{p}\cdot(\bm{r}-\bm{r}')} \quad {\rm with }
  \int_{\bm{p}}\equiv
  \int_{\mathbb{R}^{d-1}} \frac{\mathrm{d}^{d-1}p}{(2\pi)^{d-1}}\;,
\end{equation}
obeys the equation
\begin{equation}
 \label{eq:Gpzzp}
 \big[ \mathring{\sigma}\,(p_<^2-\partial_z^2)^2+p_>^2
  +\mathring{\rho}\,(p_<^2-\partial_z^2)
  +\mathring{\tau}\big]\hat{G}(\bm{p};z,z') =\delta(z-z')\;,
\end{equation}
provided the order parameter profile $\langle\bm{\phi}(\bm{x})\rangle$
vanishes (as it does in the disordered phase). This equation must be
solved subject to the boundary conditions
\begin{equation}\fl
  \label{eq:bc1pspace}
  {\Big\{\mathring{\sigma}\partial_z^3+\big[\mathring{b}
    +(\mathring{f}-\mathring{\sigma})p_<^2-  
    \mathring{\rho}\big]\partial_z+\mathring{c}_\perp
+{\mathring{\lambda}_\|} \,p_<^2
    \Big\}}\hat{G}(\bm{p};z=0,z') =0\;,
\end{equation}
\begin{equation}\fl
  \label{eq:bc2pspace}
  {\left[-\mathring{\sigma}\,\partial_z^2+\mathring{\lambda}_\perp\partial_z
  +\mathring{b}
    +(\mathring{f}+\mathring{\sigma})p_<^2\right]}\hat{G}(\bm{p};z=0,z')=0\;,
\end{equation}
and the requirement that the correct bulk propagator
$\hat{G}_{\mathrm{b}}(\bm{p},z-z')$ is obtained as $z,\,z'\to\infty$
at fixed $z-z'$. Furthermore, in order that the highest derivative,
$\partial_z^4$, produces the $\delta$-function in
equation~(\ref{eq:Gpzzp}), both $\hat{G}(\bm{p};z,z')$ and its bulk
counterpart must satisfy the jump condition
\begin{equation}
  \label{eq:jumpcond}
  \big[\mathring{\sigma}\partial_z^3
    \hat{G}(\bm{p};z,z')\big]^{z=z'+0}_{z=z'-0} =1\;.
\end{equation}
Further, $\hat{G}_(\bm{p};z,z')$ must be symmetric under exchange of
$z$ and $z'$ because $G(\bm{x},\bm{x}')$ is the inverse of a symmetric
integral kernel $A(\bm{x},\bm{x}')$ associated with the Gaussian part
$\int_{\mathfrak{V}}\rmd^dx\int_{\mathfrak{V}}\rmd^dx'\,
\bm{\phi}(\bm{x})\,A(\bm{x},\bm{x}')\,\bm{\phi}(\bm{x}')/2$ of the
Hamiltonian (including boundary terms).

To simplify our analysis, we restrict ourselves to the LP, setting
$\mathring{\rho}=\mathring{\tau}=0$. It is also convenient to set the
variable $\mathring{\sigma}$ (whose renormalized counterpart $\sigma$
changes under RG transformations) temporarily to unity. The dependence
on it can be re-introduced whenever needed by elementary dimensional
considerations. For instance, for the free propagator these lead to
the relation
\begin{equation}
  \label{eq:Gsigma}
  \hat{G}\big(\bm{p}_<,\bm{p}_>;z,z'|\mathring{\sigma}\big)
  =\mathring{\sigma}^{-1/4}\,
  \hat{G}\big(\mathring{\sigma}^{1/4}\bm{p}_<,\bm{p}_>; 
  \mathring{\sigma}^{-1/4}z,\mathring{\sigma}^{-1/4}z'|1\big)\;.
\end{equation}

The bulk propagator at the LP can be computed in a straightforward
fashion. One obtains
\begin{eqnarray}
  \label{eq:Gbpz}
  \hat{G}_{\mathrm{b}}(\bm{p},z)&=&\int_{-\infty}^\infty\frac{dk}{2\pi}\,
  \frac{\mathrm{e}^{ikz}}{(k^2+p_<^2)^2+p_>^2}\\ 
&=&
\frac{\mathrm{e}^{-\kappa_+|z|}}{4\kappa_+\kappa_-(\kappa_+^2+\kappa_-^2)}
\left[
\kappa_- \cos(\kappa_-|z|)+\kappa_+ \sin(\kappa_-|z|)
\right]\;,
\end{eqnarray}
where we have written the roots of the denominator of the required
Fourier integral as $\pm\kappa_-\pm\rmi\kappa_+$, with
\begin{equation}
\label{eq:kappapm}
\kappa_{\pm}=\frac{1}{\sqrt{2}}\,\sqrt{\sqrt{p_<^4+p_>^2}\pm
  p_<^2}\;.
\end{equation}

In terms of the linearly independent solutions
\begin{equation}
  \label{eq:Us}
  W_j(\bm{p},z)=\cases{\rme^{-\kappa_+z}\,\cos(\kappa_-z)&for $j=1$,\cr
    \rme^{-\kappa_+z}\,\sin(\kappa_- z)&for $j=2$,\cr
    \rme^{\kappa_+z}\,\cos(\kappa_-z)&for $j=3$,\cr 
    \rme^{\kappa_+z}\,\sin(\kappa_- z)&for $j=4$, }
\end{equation}
of the homogeneous counterpart of equation~(\ref{eq:Gpzzp}) with
$\mathring{\rho}=\mathring{\tau}=0$ and $\mathring{\sigma}=1$, the
free propagator of the semi-infinite system can be expressed as
\begin{equation}
  \label{eq:Ggenform}
  \hat{G}(\bm{p};z,z')=
  \theta(z-z')\sum_{j=1}^2W_j(\bm{p},z)\,V_j(\bm{p},z')\;,
   +(z\leftrightarrow z')
\end{equation}
where
\begin{equation}
  \label{eq:Vs}
  V_j(\bm{p},z')=\sum_{k=1}^4C_{jk}\,W_k(\bm{p},z').
\end{equation}
The coefficients $C_{jk}$ are chosen such that $V_{j=1,2}(\bm{p},z)$
satisfy the two boundary conditions~(\ref{eq:bc1pspace}) and
(\ref{eq:bc2pspace}) with $\mathring{\rho}=0$ and
$\mathring{\sigma}=1$, that $\hat{G}(\bm{p};z,z')$ and its first and
second derivatives with respect to $z$ are continuous at $z=z'$, and
the jump condition~(\ref{eq:jumpcond}) is fulfilled (again with
$\mathring{\sigma}=1$).

It is convenient to split off the known bulk propagator, writing
\begin{equation}
  \label{eq:Gdec}
  \hat{G}(\bm{p};z,z')= \hat{G}_{\mathrm{b}}(\bm{p};z,z') +
  \hat{G}_{\mathrm{s}}(\bm{p};z,z')\;.
\end{equation}
Here $\hat{G}_{\mathrm{s}}$, the part due to the surface, is (at
least) four times differentiable in $z$ and $z'$, and does not
contribute to the jump of $\partial_z^3\hat{G}$ at $z=z'$, which
originates entirely from $\hat{G}_{\mathrm{b}}$. Thus the jump
condition~(\ref{eq:jumpcond}) is taken care of. We can decompose
$\hat{G}_{\mathrm{b}}$ in the same way as we did for $\hat{G}$ in
equations~(\ref{eq:Ggenform}) and (\ref{eq:Vs}), with corresponding
coefficients $C_{jk}^{(\mathrm{b})}$. Evidently, only
$C_{jj'+2}^{(\mathrm{b})}$ with $j,j'=1,2$ are nonzero and can be read
off from equation~(\ref{eq:Gbpz}), but $C_{jj'}^{(\mathrm{b})}=0$.
Since $\hat{G}_{\mathrm{s}}$ must not have exponentially increasing
parts, four of the coefficients $C_{jk}$, namely
$C_{jk}=C_{jk}^{(\mathrm{b})}$ with $k=3,4$, follow immediately from
our result~(\ref{eq:Gbpz}) for $\hat{G}_{\mathrm{b}}$:
\begin{eqnarray}
  \label{eq:Cb}
  C_{13}=C_{24}=\frac{1}{4\kappa_+(\kappa_+^2 +\kappa_-^2)}=
  -\frac{\kappa_-}{\kappa_+}\,C_{14}=
 \frac{\kappa_-}{\kappa_+}\,C_{23}\;.
\end{eqnarray}
The remaining four coefficients must be determined from the two boundary
conditions~(\ref{eq:bc1pspace}) and (\ref{eq:bc2pspace}) for $V_1$ and
$V_2$.  A straightforward calculation yields
\begin{eqnarray}
C_{11}=\frac{C_{13}}{D}\left[B_++p_>^2 + 8\kappa_+^4
+L-C\right]\;,
\nonumber\\[\smallskipamount] 
C_{12}=C_{21}=\frac{C_{23}}{D}\left[B_- -p_>^2+L-C\right]\,,
\nonumber \\[\smallskipamount]
C_{22}=\frac{C_{13}}{\kappa_-^2 D}\left[B_\kappa+p_>^2\kappa_-^2-\kappa_-^2 L+ C_\kappa\right]\,,
\end{eqnarray}
where we introduced the short-hand notations
\begin{eqnarray}\fl
\mathring{b}_{p_<}\equiv\mathring{b}+\mathring{f}p_<^2\,,\quad
\mathring{c}_{p_<}\equiv\mathring{c}_\perp+\mathring{\lambda}_\|
p_<^2\,,\nonumber\\ \fl
B_+=\mathring{b}_{p_<}(\mathring{b}_{p_<}+4\kappa_+^2)\,,\quad
B_-=\mathring{b}_{p_<}(\mathring{b}_{p_<}-4\kappa_-^2)\,,\quad
B_\kappa=\mathring{b}_{p_<}\big[\mathring{b}_{p_<}(\kappa_-^2 +
2\kappa_+^2)-p_>^2\big]\,,\nonumber \\ \fl  
C=\mathring{c}_{p_<}(\mathring{\lambda}_\perp+2\kappa_+)\,,\quad
C_\kappa=\mathring{c}_{p_<}\big[2\kappa_-^2\kappa_+
-\mathring{\lambda}_\perp(\kappa_-^2+2\kappa_+^2) \big]\,,\nonumber \\
\fl L=2\mathring{\lambda}_\perp\kappa_+(\kappa_-^2+\kappa_+^2)\,,\quad
D=-B_++p_>^2+L+C\,. \label{eq:D}
\end{eqnarray}

The resulting lengthy expression for the free propagator for general
values of the surface interaction constants will not be used in the
rest of the paper. The explicit form of the free propagator that we
shall actually utilize is given in the next section, see
equation~(\ref{eq:G00prop}).

\section{Asymptotic boundary conditions at the ordinary transition}
\label{sec:BOT}

We now turn to the analysis of the ordinary transition. Let us begin
by considering first the simpler case of the ordinary transition at a
\emph{critical point} (CP), and recall some essentials of its RG
analysis. The critical behaviour at this transition is described by a
fixed point at which the Dirichlet boundary condition $\phi|_{z=0}=0$
holds. One can choose this boundary condition from the outset (already
for the bare theory) by setting the interaction constant
$\mathring{c}$ of the corresponding surface density~(\ref{eq:L1m0}) to
the value $\mathring{c}_{\mathrm{ord}}=\infty$
\cite{Die86a,Die97,DD80,DD81a}. To obtain the behaviour of $N$-point
correlation functions involving fields $\bm{\phi}$ close to the
surface, one can employ the boundary operator expansion (BOE)
\cite{Die86a,Die97,DD81a} 
\begin{equation}
  \label{eq:BOECP}
  \bm{\phi}^{\mathrm{ren}}(\bm{r},z)\mathop{\approx}\limits_{z\to 0}
  C_{\partial_n\phi}(z)\,(\partial_n\bm{\phi})^{\mathrm{ren}}(\bm{r}) \;,
\end{equation}
where $\bm{\phi}^{\mathrm{ren}}$ and $(\partial_n\phi)^{\mathrm{ren}}$
denote renormalized operators. The required independent surface
critical exponent $\beta_1^{\mathrm{ord}}$, which together with the usual bulk
critical indices yields the other surface critical exponents of this
CP ordinary transition via scaling laws, can be inferred from the
scaling dimension $\Delta[\partial_n\phi]=\beta_1^{\mathrm{ord}}/\nu$
of $\partial_n\bm{\phi}$ or the behaviour
\begin{equation}
  \label{eq:Cpartialnphi}
  C_{\partial_n\phi}(z)\sim z^{(\beta-\beta_1^{\mathrm{ord}})/\nu}
\end{equation}
of the coefficient function $C_{\partial_n\phi}$.

This CP ordinary transition must be recovered when $\rho>0$ for
appropriate choices of the surface interaction constants. It is
therefore reasonable to expect that the Dirichlet boundary condition
will prevail when $\rho\to 0$, i.e., at the fixed point describing the
LP ordinary transition. To show this and to find the required second
boundary condition, note that the surface variables with the largest
$\mu$-dimension are $\mathring{c}_\perp$ and $\mathring{b}$. At the
trivial Gaussian fixed point, their dimensionless counterparts
$c_\perp\equiv \mu^{-3/2}\mathring{c}_\perp$ and $b\equiv
\mu^{-1}\mathring{b}$ transform under scale transformations
$\mu\to\mu\ell$ as
\begin{equation}
  \label{eq:cbGauss}
  c_\perp\to \bar{c}_\perp(\ell)=\ell^{-3/2}\,c_\perp\quad\mbox{and}\quad
  b\to\bar{b}(\ell)=\ell^{-1}\,b\;.
\end{equation}
Since the ordinary transition corresponds to the case where order is
suppressed near the surface, we can take $\mathring{c}_\perp$ to be
positive. Thus $\bar{c}_\perp\to +\infty$ in the large length-scale
limit $\ell\to 0$. Further, unless fine-tuned, $\mathring{b}$ is not
expected to vanish. Equation~(\ref{eq:cbGauss}) tells us that
$\bar{b}(\ell)/\bar{c}(\ell)$ approaches zero as $\ell\to 0$. The
other coupling constants appearing in the boundary
conditions~(\ref{eq:bcperp1}), (\ref{eq:bcperp2}),
(\ref{eq:bc1pspace}) and (\ref{eq:bc2pspace}) with $\mathring{\rho}=0$
have $\mu$-dimensions smaller than both $\mathring{c}_\perp$ and
$\mathring{b}$. Hence the ratio of the associated running variables
and either $\bar{c}_\perp$ or $\mathring{b}$ also becomes small as
$\ell\to 0$. We conclude that both $\bm{\phi}|_{z=0}$ and
$\partial_n\bm{\phi}$ must become small in this limit in order for
these boundary conditions to hold. In other words, the boundary
condition~(\ref{eq:bcperp}) should apply in the large-length-scale
limit. In the following we exploit this idea by choosing them from the
outset for the bare theory.

Let us denote by $\hat{G}_{00}(\bm{p};z,z')$
the free propagator that satisfies the boundary conditions
\begin{equation}
  \label{eq:bcGDD}
  \left.\hat{G}_{00}(\bm{p};z,z')\right|_{z=0}
  =\left.\partial_z\hat{G}_{00}(\bm{p};z,z')\right|_{z=0}=0\,.
\end{equation}
It can be determined in a
straightforward manner. One gets
\begin{equation}\fl
  \label{eq:G00prop}
  \hat{G}_{00}(\bm{p};z,z')=
  \hat{G}_{\mathrm{b}}(\bm{p};z-z')-\hat{G}_{\mathrm{b}}(\bm{p};z+z')
  -\frac{\sin(\kappa_-z)\sin(\kappa_-z')}{2\kappa_-^2\,
    \kappa_+}\,\rme^{-\kappa_+(z+z')}\;.
\end{equation}

Information about correlations near the surface can again be obtained
via the BOE. Because of the boundary conditions~(\ref{eq:bcperp}), the
leading operator contributing to the BOE of $\bm{\phi}$ now is
$\partial_n^2\bm{\phi}$; we have
\begin{equation}
  \label{eq:BOELP}
  \bm{\phi}^{\mathrm{ren}}(\bm{r},z)\mathop{\approx}\limits_{z\to 0}
  C_{\partial^2_n\phi}(z)\,(\partial^2_n\bm{\phi})^{\mathrm{ren}}(\bm{r})
\end{equation}
instead of equation~(\ref{eq:BOECP}). The renormalized operator will
be defined in the next section. Suffice it here to say that its
scaling dimension
\begin{equation}
  \label{eq:Delphipp}
  \Delta[\partial_n^2\phi]=
  \beta_{\mathrm{L}1}^{(\mathrm{ord},\perp)}/\nu_{\mathrm{L}2}\;,
\end{equation}
where $\nu_{\mathrm{L}2}$ is a bulk correlation-length exponent, gives
us the surface critical exponent
$\beta_{\mathrm{L}1}^{(\mathrm{ord},\perp)}$, and that by analogy with
equation~(\ref{eq:Cpartialnphi}) we have
\begin{equation}
  \label{eq:Cpartn2phi}
   C_{\partial_n^2\phi}(z)\sim z^{(\beta_{\mathrm{L}}
     -\beta_{\mathrm{L}1}^{(\mathrm{ord},\perp)})/(\theta\nu_{\mathrm{L}2})}\;.
\end{equation}

\section{RG analysis of the ordinary transition}
\label{sec:RG}

Focusing our attention on the theory with the boundary
conditions~(\ref{eq:bcperp}), we introduce the $(N+M)$-point cumulants
\begin{equation}
  \label{eq:GNMdef}
  G^{(N,M)}_{a_1,\ldots,b_M}(\bm{x},\bm{r})\equiv\bigg\langle
    \prod_{i=1}^N\phi_{a_i}(\bm{x}_i)\,
    \prod_{j=1}^M\partial_n^2\phi_{b_j}(\bm{r}_j)
    \bigg\rangle^{\mathrm{cum}}.
\end{equation}
To regularize their ultraviolet (uv) singularities, we employ
dimensional regularization. Apart from bulk uv singularities known
from studies of bulk critical behaviour \cite{DS00a,SD01,DSZ03}, they
have (primitive) uv surface singularities originating from the surface
part of the free propagator, i.e.\ the last two terms on the
right-hand side (rhs) of equation~(\ref{eq:G00prop}). We use the
reparametrization convention of references~\cite{DSZ03,DGR03,DR04} to
absorb the uv bulk singularities, introducing renormalized quantities
and renormalization factors via
\begin{eqnarray}
  \label{eq:bulkreps}
  \bm{\phi}=Z_\phi^{1/2}\bm{\phi}^{\mathrm{ren}}\,,\quad
  \mathring{\sigma}=Z_\sigma\,\sigma\,,\quad
  \mathring{u}\mathring{\sigma}^{-m/4}= F_{m,\epsilon}\,\mu^\epsilon
  Z_u u\,,  \\
\mathring{\tau}-\mathring{\tau}_{\mathrm{LP}}=
  \mu^2Z_\tau\big[\tau+A_\tau\,\rho^2\big]\,, \quad (\mathring{\rho}
  -\mathring{\rho}_{\mathrm{LP}})\mathring{\sigma}^{-1/2}= \mu Z_\rho\rho\,,
\end{eqnarray}
with
\begin{equation}
  \label{eq:Fmeps}
  F_{m,\epsilon}=\frac{\Gamma(1+\epsilon/2)\, \Gamma^2(1-\epsilon/2)\,
    \Gamma(m/4)}{(4\pi)^{(8+m-2\epsilon)/4}\,\Gamma(2-\epsilon)\,
    \Gamma(m/2)}\;,
\end{equation}
where we have explicitly indicated the dependence on
$\mathring{\sigma}$. Two-loop results for the renormalization factors
$Z_{\phi}$, $Z_\sigma$, $Z_\rho$ and $Z_\tau$ can be found in
reference~\cite{SD01}. The renormalization function $A_\tau$ was
computed to one-loop order in reference~\cite{DGR03}; the result will
not be needed in the following.

In order to absorb the primitive uv surface singularities, additional
counter-terms with support on the surface are needed. We introduce a
renormalization factor $Z_2$ and renormalized surface operator
$(\partial^2_n\bm{\phi})^{\mathrm{ren}}$ via
\begin{equation}
  \label{eq:pn2phiren}
  \partial^2_n\bm{\phi}=\big[Z_{2}\,Z_\phi\big]^{1/2}\,
  (\partial^2_n\bm{\phi})^{\mathrm{ren}}
  =Z_2^{1/2}\partial_n^2\bm{\phi}^{\mathrm{ren}} \;.
\end{equation}
Power counting shows that besides the surface counter-term resulting
from $Z_2$, $G^{(0,2)}$ requires an additive renormalization. Let us
write the source term of the action needed to generate insertions of
the bare surface operator $\partial_n^2\bm{\phi}$ as
$\int_{\mathfrak{B}}\rmd^{d-1}r\;\bm{J}_2(\bm{r})\cdot
\partial_n^2\bm{\phi}(\bm{r})$. Then the $\bm{J}_2$-dependent part of
the renormalized action can be written as
\begin{equation}
  \label{eq:AJ2}
 \int_{\mathfrak{B}}\rmd^{d-1}r\left[
   Z_2^{-1/2}\, \bm{J}_2\cdot
   \partial_n^2\bm{\phi}^{\mathrm{ren}} +\mu^{1/2}\sigma^{-5/4}
   S_2\,\bm{J}_2^2\right] \;.
\end{equation}

\begin{figure}[htb]
  \centering
  \includegraphics[width=10em]{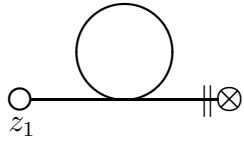}
  \caption{One-loop graph of $\hat{G}^{(1,1)}(\bm{p},z_1)$. The stroke
    indicates the derivative $\partial_z$. Crossed and open circles
    indicate points on and off the surface, respectively. Thus the
    crossed circle in conjunction with the double stroke on the right
    represents the surface operator $\partial^2_n\bm{\phi}$.}
  \label{fig:onelg}
\end{figure}

This requires some explanation. The counter-term involving
$Z_2$ absorbs uv singularities $\sim \ln\Lambda$; as usual, the latter
manifest themselves as poles at $\epsilon=0$ in the dimensional
regularization scheme we prefer to employ. That such poles indeed
occur, can be seen from the one-loop graph of $G^{(1,1)}$ shown in
figure~\ref{fig:onelg}. %
Its closed loop involves the distribution
\begin{equation}
  \label{eq:G00}
  G_{00}(\bm{x},\bm{x})=
  \mathring{\sigma}^{-1/4}\,\mu^{-3/2}\,\check{z}^{-4+2\epsilon}
  \,g_\epsilon(m)\,F_{m,\epsilon}
  \;,
\end{equation}
in which $\check{z}\equiv\mathring{\sigma}^{-1/4}\mu^{1/2}z$ denotes
the dimensionless counterpart of $z$, while $g_\epsilon(m)$ is a
number given by
\begin{equation}
  \label{eq:gepsm}
  g_\epsilon(m)=F_{m,\epsilon}^{-1}\int_{\bm{p}}
  \hat{G}_{00}(\bm{p};1,1|\mathring{\sigma}=1) \;.
\end{equation}
Since $\check{z}\geq 0$, the power $\check{z}^{-4+2\epsilon}$
corresponds to the generalized function $\check{z}_+^{-4+2\epsilon}$
of reference~\cite{GS64}. Upon substituting its well-known Laurent
expansion \cite{GS64} (see also the appendix of
reference~\cite{Die86a}), we obtain
 \begin{equation}
  \label{eq:G00Laurent}
  G_{00}(\bm{x},\bm{x})/F_{m,\epsilon}=
  \mathring{\sigma}^{-1/4}\,\mu^{-3/2}\,  g_0(m)\, \frac{-1}{6}\,
  \frac{\delta^{\prime\prime\prime}(\check{z})}{2\epsilon}
  +\Or(\epsilon^0)\;.
\end{equation}
The quantity $g_\epsilon(m)$ is computed in the appendix for general
values of $m$. Our subsequent results involve its value at
$\epsilon=0$, which is
\begin{eqnarray}
 \label{eq:g0m}
\fl
g_0(m)=2-m-\frac{3}{4-m}\bigg\{2-m+\frac{24}{m+2}
-\frac{(2\pi)^{1/2}\,(2m-5)(m-1)\,\Gamma(m/2)}{8\,\Gamma[(m+1)/2]}\nonumber\\
\strut +\frac{1}{m}\; _2F_1[2,(m-1)/2;(m+2)/2;-1]
\bigg\}\,,
\end{eqnarray}
and reduces to
\begin{equation}
  \label{eq:g01}
  g_0(1)=-9
\end{equation}
in the uniaxial case $m=1$.

In accordance with its indicated $\mu$-dimension, the counter-term
$\propto S_2$ diverges $\sim \sqrt{\Lambda}$. Its is analogous to the
surface counter-term $C_\infty\sim \Lambda$ in equation~(3.135) of
reference~\cite{Die86a} needed in the CP case to renormalize the
corresponding correlation functions $\langle\phi_{a_1}
\cdots\phi_{a_N}\partial_n\phi_{b_a} \cdots\partial_n\phi_{b_M}
\rangle$ in the cutoff-regularized theory. Just as ${C_\infty\sim
  \Lambda}$, the counter-term $\propto S_2$ vanishes in the
dimensionally regularized theory. We therefore do not consider it any
further.

Suppressing the tensorial indices $a_1,\ldots,b_M$, we denote the
renormalized counterparts of the cumulants~(\ref{eq:GNMdef}) as
\begin{equation}
  \label{eq:GNMren}
  G^{(N,M)}_{\mathrm{ren}}=Z_\phi^{-(N+M)/2}Z_2^{-M/2}\,G^{(N,M)}\;. 
\end{equation}
A standard way of reasoning yields their RG equations
\begin{equation}
  \label{eq:RGE}
  \bigg[\mu\partial_\mu+\sum_{\wp=u,\sigma,\rho,\tau}
  \beta_\wp\partial_\wp +\frac{N+M}{2}\,\eta_\phi 
  +\frac{M}{2}\,\eta_2\bigg]G^{(N,M)}_{\mathrm{ren}} =0\;,
\end{equation}
where $\beta_{\wp}$ are bulk $\beta$-functions in the notation of
references~\cite{DGR03,DR04}. Both $\beta_\rho$ as well as
$\beta_\tau$ vanish at the LP and will not be needed in the rest of
the paper. The functions $\beta_u(\epsilon,u)$ and
$\eta_\sigma(u)\equiv -\beta_\sigma/\sigma$ are known to order $u^3$
from reference \cite{SD01}.

As a consequence of these RG equations, we have for the coefficient
function $C_{\partial^2_n\phi}(z)$,
\begin{equation}
  \label{eq:RGEC}
   \bigg[\mu\partial_\mu+\sum_{\wp=u,\sigma,\rho,\tau} \beta_\wp
   \partial_\wp -\frac{\eta_2}{2}\bigg]C_{\partial^2_n\phi}(z)=0\;.
\end{equation}
Solving this at the LP $\rho=\tau=0$ gives
\begin{equation}
  \label{eq:Czsol}
  C_{\partial^2_n\phi}(z)\sim \big(\mu\sigma^{-1/2}
    \big)^{\eta_2^*/(4\theta)}\, z^{(\eta^*_2+4\theta)/(2\theta)} \;,
\end{equation}
where $\eta_2^*=\eta_2(u^*)$ means the value of $\eta_2$ at the
infrared-stable fixed point $u^*=\Or(\epsilon)$. The result can be
combined with equation~(\ref{eq:Cpartn2phi}) to conclude that
\begin{eqnarray}
  \label{eq:beta1Lgr}
  \beta_{\mathrm{L}1}^{(\mathrm{ord},\perp)}
&=&\big(d-m-2+\eta_{\mathrm{L}2} +m\theta
    +\eta_2^*+4\theta\big)\nu_{\mathrm{L}2}/2\nonumber\\
&=& \beta_{\mathrm{L}}+
  \big(4\theta+\eta_2^*\big)\nu_{\mathrm{L}2}/2\;.
\end{eqnarray}
Here $\beta_{\mathrm{L}}=\nu_{\mathrm{L}2}[d-2+\eta_{\mathrm{L}2}
+m(\theta-1)]$ is the usual bulk magnetization exponent. 

The scaling behaviour of other surface quantities can be derived along
similar lines by exploiting the RG equations~(\ref{eq:RGE}) in
conjunction with the BOE~(\ref{eq:BOELP}). In this manner one can
determine how the associated surface critical exponents can be
expressed in terms of four independent bulk critical exponents, such
as $\eta_{\mathrm{L}2}$, $\theta$, $\nu_{\mathrm{L}2}$ and $\varphi$,
and the surface magnetization index
$\beta_{\mathrm{L}1}^{(\mathrm{ord},\perp)}$.

As an example, let us consider the surface susceptibility
$\chi_{11}(\bm{p}) = {\hat{G}^{(2,0)}(\bm{p};z=0,z'=0)}$, the Fourier
$\bm{p}$-transform of the response of
$\langle\bm{\phi}|_{z=0}\rangle$ with respect to a surface magnetic
field $\bm{h}_1$. To characterize the asymptotic low-momentum
behaviour of this quantity at the LP, we introduce the surface
exponent $\eta_\|$ by analogy with the CP case via
\begin{equation}
  \label{eq:chi11}
  \chi_{11}^{(\mathrm{sing})}(\bm{p}) \mathop{\sim}\limits_{\bm{p}\to\bm{0}}
  \cases{{p_<}^{(\eta_\|-1)/\theta}&for $p_>=0$\;,
    \cr {p_>}^{\eta_\|-1}& for $p_<=0$\;,}
\end{equation}
where ``sing'' means singular part. (At the ordinary transition
considered here, $\chi_{11}$ does not diverge but approaches a finite
value.)

To identify $\eta_\|$, we apply the BOE~(\ref{eq:BOELP}) to both
external legs of $\hat{G}^{(2,0)}(\bm{p},z,z')$, and exploit the
behaviour~(\ref{eq:Czsol}) of the coefficient function together with
the scaling forms of $G^{(2,0)}$ and $G^{(0,2)}$ implied by their RG
equations. This gives, for example,
\begin{equation}
  \label{eq:chi11sing}
  \hat{G}^{(2,0)}_{\mathrm{ren}}(\bm{p};z,z')\mathop{\sim}\limits_{z,z'\to 0} 
  {p_>}^{-2+\eta_{\mathrm{L}2}+\theta} \,
  \big[\big(zp_>^\theta)(z'p_>^\theta\big)
  \big]^{(\eta_2^*+4\theta)/(2\theta)}\;,
\end{equation}
which tells us that
\begin{equation}
  \label{eq:etaparscal}
  \eta_\|^{(\mathrm{ord},\perp)}=5\theta-1+\eta_{\mathrm{L}2}+\eta_2^*\;.
\end{equation}
Other surface exponents, such as the surface susceptibility exponents
$\gamma_{11}$ and $\gamma_1$ can be expressed in a similar manner in
terms of $\beta_{\mathrm{L}1}^{(\mathrm{ord},\perp)}$ and bulk
  exponents. In particular, the usual scaling law
  $\gamma_{11}=\nu_{\mathrm{L}2}(1-\eta_\|)$ for the surface
  susceptibility exponent $\gamma_{11}$ is found to remain valid, as
  suspected by Binder and Frisch \cite{BF99}.

Next, we turn to the computation of the still unknown RG~function
$Z_2$. Using the result~(\ref{eq:G00Laurent}), one can determine it to
one-loop order in a straightforward fashion.  Application of the
distribution $-\delta'''(z)$ to the external legs of
$\hat{G}^{(0,2)}(\bm{p})$ and $\hat{G}^{(1,1)}(\bm{p};z_1)$ gives
\begin{equation}
  \Big(-\delta^{\prime\prime\prime}(z),\Big[\hat{G}_{00}
 \overleftarrow{\partial}_n^2\Big]^2\Big)=6\,
 \partial_n^2\hat{G}_{00}\overleftarrow{\partial}_n^2 \;.
\end{equation}
and
\begin{equation}
    \left(-\delta^{\prime\prime\prime}(z), \hat{G}_{00}(\bm{p};z_1,z)\,
      \Big[\hat{G}_{00} \overleftarrow{\partial}_n^2\Big](\bm{p};z,0)
    \right)=3\, \Big[\hat{G}_{00}\overleftarrow{\partial}_n^2\Big](z_1,0)\;,
\end{equation}
respectively. The implied poles of both
$\hat{G}_{\mathrm{ren}}^{(0,2)}(\bm{p})$ and
$\hat{G}^{(1,1)}_{\mathrm{ren}}(\bm{p};z_1)$ get cancelled if we choose
\begin{equation}
  \label{eq:Z2res}
  Z_2=1-g_0(m)\,\frac{n+2}{3}\,\frac{u}{4\epsilon}+\Or(u^2)\;,
\end{equation}
where we utilized the fact that $Z_\phi=1+\Or(u^2)$. The derivative
$-u\partial_u$ of this function's residuum at $\epsilon=0$ gives us
the RG~function $\eta_2$. Substituting for $u$ its fixed-point value
$u^*=6\epsilon/(n+8)+\Or(\epsilon^2)$ then yields
\begin{equation}
  \label{eq:eta2}
  \eta_2^*=g_0(m)\,\frac{n+2}{12}\,u^*+\Or\big[(u^*)^2\big]
  =g_0(m)\,\frac{n+2}{n+8}\,\frac{\epsilon}{2}+\Or(\epsilon^2)\;.
\end{equation}
The result can be combined with the known $\epsilon$~expansions of the
bulk exponents in equation~(\ref{eq:beta1Lgr}) to obtain
\begin{eqnarray}
  \label{eq:betaL1epsex}
   \beta_{\mathrm{L}1}^{(\mathrm{ord},\perp)}&=&
   (\nu_{\mathrm{L}2}/2)[4- \epsilon+\eta_2^*]
   +\Or(\epsilon^2)\nonumber\\&=&1+\frac{\epsilon}{4(n+8)} 
\left[n-4+\frac{n+2}{2}\,g_m(0)\right]+\Or(\epsilon^2)\,.
\end{eqnarray}

In the uniaxial scalar case $m=n=1$ of the ANNNI model, this
simplifies to
\begin{equation}
  \label{eq:beta1ANNNI}
  \beta_{\mathrm{L}1}^{(\mathrm{ord},\perp)}=1-\frac{11}{24}\,\epsilon
  +\Or(\epsilon^2)=\left[1+\frac{11}{24}\,\epsilon
    +\Or(\epsilon^2)\right]^{-1}\;.
\end{equation}
Setting $\epsilon=3/2$ in the first and second expressions on the rhs
(i.e., in the ``direct series'' and the [0/1] Pad{\'e} approximant)
yields the $d=3$ estimates
$\beta_{\mathrm{L}1}^{(\mathrm{ord},\perp)}\simeq 0.31$ and $0.59$,
respectively, which is to be compared with Pleimling's Monte Carlo
result $0.62(1)$ \cite{Ple02}. 
Though these numbers are encouraging, our present knowledge of the
series~(\ref{eq:beta1ANNNI}) to just first order in $\epsilon$ clearly
is insufficient to produce estimates that are competitive in accuracy
with this Monte Carlo value. We therefore refrain from giving further
extrapolated values for other surface exponents. Experience with the
bulk case \cite{SD01} suggests that much better field-theoretic
estimates should become possible once $\eta_2^*$ is known to
$\Or(\epsilon^2)$. In view of the simplifications entailed by the
approach developed here, such a two-loop calculation should not be too
difficult.

\section{Concluding remarks}
\label{sec:concl}

In this paper we have extended previous field-theoretic work on
boundary critical behaviour at $m$-axial LPs
\cite{DRG03,DGR03,DR04,Die05} by studying the critical behaviour at
the ordinary transition of a semi-infinite system that is bounded by a
surface perpendicular to an $\alpha$-direction. This geometry was the
one considered in the earliest investigations \cite{Gum86,BF99} of
boundary critical behaviour at LPs. However, to construct an
appropriate minimal field-theory model and analyze it in a systematic
fashion by means of modern RG method below the upper critical
dimension turned out to be quite a challenge, mainly because
significantly more potentially relevant (short-range) surface
contributions must be considered than in the simpler case of where the
surface is perpendicular to $\beta$-direction.

Taking up a suggestion made earlier \cite{Die05}, we found a way to
by-pass the enormous technical difficulties one is faced with when
having to carry along many surface interaction constants. The basic
idea is to choose the boundary conditions~(\ref{eq:bcperp}) the theory
is expected to satisfy on long length-scales at the ordinary
transition from the outset, showing that they \emph{correspond to a
  fixed point} of the RG. While we have not investigated deviations
from this fixed point associated with modification of these boundary
conditions (e.g., finite values of the surface couplings
$\mathring{c}_\perp$ and $\mathring{b}$), our results are completely
consistent with the physically reasonable expectation that these
boundary conditions are associated with a stable fixed point of the
RG.

The benefit of our procedure is twofold: (i) Used in conjunction with
the BOE, the RG equations we obtained enabled us to derive scaling
laws relating the surface critical exponents at this transition to
bulk exponents and a single independent surface index, such as
$\beta_{\mathrm{L}1}^{(\mathrm{ord},\perp)}$. (ii) Owing to the gain
in technical simplification, an explicit one-loop calculation could be
performed to obtain the $\epsilon$ expansion of
$\beta_{\mathrm{L}1}^{(\mathrm{ord},\perp)}$ and related surface
exponents to first order. Of course, one cannot expect to get precise
numerical estimates for the critical exponents at $d=3$ from such an
$\Or(\epsilon)$ result. Nevertheless, the extrapolated values we
obtained for $\beta_{\mathrm{L}1}^{(\mathrm{ord},\perp)}$ in the case
of the three-dimensional semi-infinite ANNNI model by direct
extrapolation to $d=3$ and via a [0/1] Pad{\'e} approximant agree
reasonably well with recent Monte Carlo results \cite{Ple02} (albeit
different numbers could be obtained if other scaling-law expressions
for $\beta_{\mathrm{L}1}^{(\mathrm{ord},\perp)}$ were employed in
conjunction with the best estimates of reference~\cite{SD01} for the
required bulk exponents). More reliable field-theoretic estimates
should be possible on the basis of $\Or(\epsilon)$ results. The
required two-loop calculation appears to be quite manageable.

A more difficult task is to perform an analogous two-loop RG analysis
of the corresponding LP special transition. Since this requires the
identification of the associated fixed point in the space of surface
coupling  constants, one cannot avoid retaining the dependence on these
variables (see the analysis  of the special transition for the case of
parallel surface orientation \cite{DR04}, for comparison).

From a more general perspective, the present investigation is, to our
knowledge, the first one dealing beyond the classical level with a
field theory subject to the two boundary conditions~(\ref{eq:bcperp}).
Both our way of investigating this field theory as well as our result
that this boundary condition correspond to a stable fixed point of the
RG might have applications in other contexts.

\ack One of us (HWD) would like to thank Royce Zia and Beate
Schmittmann for their hospitality at the Physics Department of
Virginia Tech, where part of this research was carried out.
MASh and PVP thank H W Diehl for his hospitality
during their visits at the Physics Department of the Universit{\"a}t
Duisburg-Essen, Campus Essen, where part of their work was performed.

The work of HWD and MASh was supported in part by the grant Di-378/3 from
the Deutsche Forschungsgemeinschaft (DFG). PVP's work and visit at
Essen was made possible by a fellowship within the framework of the
German-Russian program ``Michael Lomonossov'', provided in parts by
DAAD and the Russian government. We gratefully acknowledge the support
of all funding agencies.

\appendix
\section{Calculation of one-loop integral}
\label{sec:app}

Here we compute the number $g_\epsilon(m)$ introduced in
equation~(\ref{eq:gepsm}).

Let us denote the bulk propagator in position space as
$G_{\mathrm{b}}(\bm{x}-\bm{x}')$. Taking into account that
$G_{\mathrm{b}}(\bm{0})$ vanishes at the LP in dimensional
regularization, we obtain from equations~(\ref{eq:G00}) and
(\ref{eq:gepsm}) the result
\begin{equation}
  \label{eq:Gzz}\fl
g_\epsilon(m)F_{m,\epsilon}=G_{00}(\bm{x},\bm{x}|\mathring{\sigma}=1)|_{z=1}= 
  -G_{\mathrm{b}}(\bm r=0,2z|1)|_{z=1} +K_{d-m}\,K_{m-1}\,I_\epsilon(m)\;.
\end{equation}
The first term on the rhs follows from equations~(7), (A5)
and (A6) of reference~\cite{SD01}. It is given by
\begin{equation}
\label{eq:Gbxx}
G_{\mathrm{b}}(\bm 0,2|1)= 2^{-1-m} \pi^{-(6+m-2\epsilon)/4}\,
\frac{\Gamma(2-\epsilon)}{\Gamma[(m-2+2\epsilon)/4)]}\;. 
\end{equation}
In the second contribution we have split off the factors originating
from the angular integrations, with
\begin{equation}
  \label{eq:Kd}
  K_d\equiv 2\,(4\pi)^{-d/2}/\Gamma(d/2)\;.
\end{equation}
In the remaining integral we make the changes of variables $p_>\to
P=p_>/p_<^2$ and $p_<\to p=p_</\sqrt{2}$ to obtain
\begin{eqnarray}
  \label{eq:Iepsm}
  I_\epsilon(m)&=&-\int_0^\infty {\rmd}p_>\, p_>^{d-m-1}
  \int_0^\infty {\rmd}p_<\, p_<^{m-2} \,
  \frac{\sin^2(\kappa_-)}{2\kappa_-^2\, 
    \kappa_+}\,\rme^{-2\kappa_+} \nonumber\\ 
&=&- 2^\beta{\int_0^\infty} {\rmd}P\,
P^\nu\int_0^\infty {\rmd}p\, p^{\alpha-1} 
\,\frac{\sin^2(p k_-)}{k_-}\,\rme^{-2 p k_+}
\end{eqnarray}
with
\begin{equation}
\alpha=2d-4-m=4-2\epsilon\,,\quad
\beta=\frac{\alpha+1}{2}\,, \quad
\nu=\frac{\alpha-m}{2}\,.
\end{equation}
We have introduced
\begin{equation}
k_\pm\equiv\left( \sqrt{1+P^2}\pm 1 \right)^{1/2}\,,
\end{equation}
for which
\begin{equation}
k_- k_+ = P\,,\quad k_+^2 - k_-^2 = 2\,,\quad (k_+ - \rmi k_-)^2
=2\,(1-\rmi P)\,.
\end{equation}

Upon integrating by parts with respect to $p$ and using
$\sin^2x=(1-\cos{2x})/2$, we arrive at
\begin{eqnarray}
I_\epsilon(m) &=&
 \frac{2^\beta}{\alpha}\int_0^\infty {\rmd}P \,P^\nu
\bigg[
\int_0^\infty {\rmd}p\, p^\alpha \sin(2 p k_-)\,\rme^{-2 p k_+}
\nonumber\\ &&\strut
-\frac{2k_+}{k_-}\,
\int_0^\infty {\rmd}p\, p^\alpha\,
\sin^2(p k_-)\,\rme^{-2 p k_+}
\bigg]\nonumber\\ &=&
\frac{2^\beta}{\alpha}
\Bigg[
-\int_0^\infty {\rmd}P \,P^{\nu-1} k_+^2\, I_p^{(0)}+\int_0^\infty
{\rmd}P\, P^\nu I_p^{(1)} \nonumber\\ &&\strut
+\int_0^\infty {\rmd}P \,P^{\nu-1} \Big( \sqrt{1+P^2} + 1 \Big)
I_p^{(2)} \Bigg]\;
\end{eqnarray}
with 
\begin{equation}
  I_p^{(0)}=\int_0^\infty {\rmd}p\, p^\alpha\,\rme^{-2 p k_+}
  =2^{-2\beta}\Gamma(2\beta)\,k_+^{-2\beta}\,,
\end{equation}
\begin{equation}\fl
  I_p^{(1)}=\int_0^\infty {\rmd}p\, p^\alpha
  \sin(2 p k_-)\,\rme^{-2 p k_+}
  =2^{-3\beta}\,\frac{\Gamma(2\beta)}{2{\rmi}}
  \left[(1-{\rmi}P)^{-\beta}-(1+{\rmi}P)^{-\beta}
  \right]
\end{equation}
and
\begin{equation}\fl
  I_p^{(2)}=\int_0^\infty {\rmd}p\, p^\alpha
  \cos(2 p k_-)\,\rme^{-2 p k_+}=
  2^{-3\beta}\,\frac{\Gamma(2\beta)}{2}
  \left[(1-{\rmi}P)^{-\beta}+(1+{\rmi}P)^{-\beta}\right]\;.
\end{equation}

The remaining $p$-integrations can be conveniently performed by means
of Maple.\footnote{Maple 8, a product of Waterloo Maple Inc.}  After
some rearrangements the result becomes
\begin{eqnarray}\fl
\label{eq:Iepsmres}
I_\epsilon(m) = - 2^{2\epsilon-5}\,\Gamma\Big(\frac{m-1}{2}
\Big)\,\Gamma(4-2\epsilon) 
\Bigg[
\frac{\Gamma(2-m/2-\epsilon)}{\Gamma(3/2-\epsilon)}\,
\cos\Big(\pi\frac{m+2\epsilon}{4}\Big)\nonumber\\ 
\lo \strut   + (3-2\epsilon)\,2^{1-m/2-\epsilon}\,
\frac{\Gamma(1-m/4-\epsilon/2)}{\Gamma(3/2+m/4-\epsilon/2)}
\nonumber\\ 
\lo \strut  -\frac{1}{2\sqrt\pi}\;\Gamma\left(-\frac{m}{2}\right)
\,\cos\Big(\pi\frac{m-2\epsilon}{4}\Big)
\,_2\!F_1\bigg(2-\epsilon,\frac{m-2}{2};\frac{m+2}{2};-1\bigg)\nonumber\\
\lo \strut   +\frac{m-1}{2}\,\frac{\Gamma(2-m/2-\epsilon)\,
  \Gamma(m/2)}{\Gamma(2-\epsilon)\,\Gamma(m/2+1/2)}\,
\cos\Big(\pi\frac{m+2\epsilon}{4}\Big)
\times\nonumber\\ 
\lo \strut   \times _2\!F_1\bigg(2-\frac{m}{2}-\epsilon,-\frac{1}{2};1
-\frac{m}{2};-1\bigg)\Bigg].
\end{eqnarray}

For the uniaxial case $m=1$, the required integration is simple, yielding
\begin{eqnarray}
I_\epsilon(1) &=&-\sqrt 2 
\int_0^\infty {\rmd}p\, p^{1-\epsilon}\,
\sin^2\big(\sqrt{p/2}\big)\;\rme^{-\sqrt
  {2p}}\nonumber \\ &=&2^{\epsilon-7/2}\,\Gamma(4-2\epsilon)\,
[4+2^\epsilon\cos(\pi\epsilon/2) ]
\end{eqnarray}
in conformity with equation~(\ref{eq:Iepsmres}).

Upon substituting the above expression~(\ref{eq:Iepsmres}) into
equation~(\ref{eq:Gzz}) and using equation~(\ref{eq:G00}) together
with the definition~(\ref{eq:Fmeps}) of $F_{m,\epsilon}$, one obtains
the result for $g_0(m)$ given in equation~(\ref{eq:g0m}).


\section*{References}

\begin{thebibliography}{10}

\bibitem{Bin83}
K. Binder,  in {\em Phase Transitions and Critical Phenomena}, edited by C.
  Domb and J.~L. Lebowitz (Academic, London, 1983), Vol.~8, pp.\ 1--144.

\bibitem{Die86a}
H.~W. Diehl,  in {\em Phase Transitions and Critical Phenomena}, edited by C.
  Domb and J.~L. Lebowitz (Academic, London, 1986), Vol.~10, pp.\ 75--267.

\bibitem{Die97}
H.~W. Diehl, Int.\ J.\ Mod.\ Phys.\ B {\bf 11},  3503  (1997),
  cond-mat/9610143.

\bibitem{Ple04}
M. Pleimling, J. Phys. A {\bf 37},  R79  (2004).

\bibitem{Hor80}
R.~M. Hornreich, J. Magn. Magn. Mater. {\bf 15--18},  387  (1980).

\bibitem{Sel92}
W. Selke,  in {\em Phase Transitions and Critical Phenomena}, edited by C. Domb
  and J.~L. Lebowitz (Academic, London, 1992), Vol.~15, pp.\ 1--72.

\bibitem{Die02}
H.~W. Diehl, Acta physica slovaca {\bf 52},  271  (2002), proc.\ of the 5th
  International Conference ``Renormalization Group 2002'', Tatranska Strba,
  High Tatra Mountains, Slovakia, March 10--16, 2002; cond-mat/0205284.

\bibitem{HLS75a}
R.~M. Hornreich, M. Luban, and S. Shtrikman, Phys. Rev. Lett. {\bf 35},  1678
  (1975).

\bibitem{SG78}
J. Sak and G.~S. Grest, Phys. Rev. B {\bf 17},  3602  (1978).

\bibitem{MC98}
C. Mergulh{\~a}o, Jr. and C.~E.~I. Carneiro, Phys. Rev. B {\bf 58},  6047
  (1998).

\bibitem{MC99}
C. Mergulh{\~a}o, Jr. and C.~E.~I. Carneiro, Phys. Rev. B {\bf 59},  13 954
  (1999).

\bibitem{DS00a}
H.~W. Diehl and M. Shpot, Phys.\ Rev.\ B {\bf 62},  12 338  (2000),
  cond-mat/0006007.

\bibitem{SD01}
M. Shpot and H.~W. Diehl, Nucl. Phys. B {\bf 612},  340  (2001),
  cond-mat/0106105.

\bibitem{DSZ03}
H.~W. Diehl, M. Shpot, and R.~K.~P. Zia, Phys. Rev. B {\bf 68},  224415
  (2003), cond-mat/0307355.

\bibitem{DRG03}
H.~W. Diehl, S. Rutkevich, and A. Gerwinski, J. Phys. A {\bf 36},  L243
  (2003).

\bibitem{DGR03}
H.~W. Diehl, A. Gerwinski, and S. Rutkevich, Phys.\ Rev.\ B {\bf 68},  224428
  (2003), cond-mat/0308483.

\bibitem{DR04}
H.~W. Diehl and S. Rutkevich, J. Phys. A {\bf 37},  8575  (2004),
  cond-mat/0406216.

\bibitem{FKB00}
H.~L. Frisch, J.~C. Kimball, and K. Binder, J. Phys.: Condens. Matter {\bf 12},
   29  (2000).

\bibitem{Ple02}
M. Pleimling, Phys. Rev. B {\bf 65},  184406  (2002).

\bibitem{Gum86}
G. Gumbs, Phys. Rev. B {\bf 33},  6500  (1986).

\bibitem{BF99}
K. Binder and H.~L. Frisch, Eur. Phys. J. B {\bf 10},  71  (1999).

\bibitem{Die05}
H.~W. Diehl, Pramana---Journal of Physics {\bf 64},  803  (2005), proceedings
  of the 22nd International Conference on Statistical Physics ({STATPHYS 22})
  of the International Union of Pure and Applied Physics ({IUPAP}), 4--9 July
  2004, Bangalore, India; cond-mat/0407352.

\bibitem{KD92a}
M. Krech and S. Dietrich, Phys. Rev. Lett. {\bf 66},  345  (1992), [Erratum
  {\bf 67}, 1055 (1992)].

\bibitem{MO93a}
D.~M. McAvity and H. Osborn, Nucl.\ Phys.\ B {\bf 394},  728  (1993).

\bibitem{Cha70}
S. Chandrasekhar, {\em Hydrodynamic and hydromagnetic stability}, 3rd ed.
  (Oxford University Press, Oxford, 1970).

\bibitem{DD80}
H.~W. Diehl and S. Dietrich, Phys.\ Lett. {\bf 80A},  408  (1980).

\bibitem{DD81a}
H.~W. Diehl and S. Dietrich, Z.\ Phys.\ B {\bf 42},  65  (1981), erratum: {\bf
  43}, 281 (1981).

\bibitem{GS64}
I.~M. Gel'fand and G.~E. Shilov,  in {\em Generalized Functions} (Academic, New
  York and London, 1964), Vol.~1, pp.\ 1--423.

\end{thebibliography}
\end{document}